\documentclass[a4paper]{article}

\usepackage{INTERSPEECH2020}
\usepackage{multirow}
\usepackage{cite}
\title{Text-Independent Speaker Verification with Dual Attention Network}
\name{Jingyu Li, Tan Lee}
\address{Department of Electronic Engineering, The Chinese University of Hong Kong, Hong Kong}
\email{lijingyu0125@link.cuhk.edu.hk, tanlee@ee.cuhk.edu.hk}

\begin{document}

\maketitle
\begin{abstract}
This paper presents a novel design of attention model for text-independent speaker verification. The model takes a pair of input utterances and generates an utterance-level embedding to represent speaker-specific characteristics in each utterance. The input utterances are expected to have highly similar embeddings if they are from the same speaker. The proposed attention model consists of a self-attention module and a mutual attention module, which jointly contributes to the generation of the utterance-level embedding. The self-attention weights are computed from the utterance itself while the mutual-attention weights are computed with the involvement of the other utterance in the input pairs. As a result, each utterance is represented by a self-attention weighted embedding and a mutual-attention weighted embedding. The similarity between the embeddings is measured by a cosine distance score and a binary classifier output score. The whole model, named Dual Attention Network, is trained end-to-end on Voxceleb database. The evaluation results on Voxceleb 1 test set show that the Dual Attention Network significantly outperforms the baseline systems. The best result yields an equal error rate of $1.6\%$.
 
\end{abstract}

\section{Introduction}
%Task introduction and background.
Speaker verification (SV) refers to the process of determining whether an input speech utterance is from a claimed speaker. If the claimed speaker is represented by a reference utterance, the task of SV is essentially to determine whether the two utterances are from the same person or not. In text-dependent SV, all utterances are required to contain the same speech content \cite{variani2014deep}. Whilst in text-independent SV, the spoken content is unrestricted \cite{campbell1997speaker}. The present study is focused on text-independent SV with a pair of input utterances.

%Baseline models: UBM, i-vectors,
% I-vector based methods represent a main-stream approach to SV \cite{dehak2010front} . In this approach, a high-dimensional super-vector derived from Gaussian Mixture Model-Universal Background Model (GMM-UBM) is used to represent statistical information about target speakers \cite{reynolds2000speaker}. The super-vector is projected into the total variability space via a trainable low-rank matrix. The projected vector, referred as i-vector, is a low-dimensional speaker-dependent representation. The similarity between speakers (or utterances) can be measured by, for example, the cosine distance between the respective i-vectors.

%DNN development, and x-vectors.
In recent years, embeddings learned by deep neural network (DNN) are widely applied to both text-dependent and text-independent SV \cite{variani2014deep,snyder2018x}. In a typical DNN pipeline of embedding generation, the input speech utterance is first converted into frame-level acoustic representations, e.g., log Mel-filterbank (FBank) or Mel frequency cepstral coefficients (MFCC). The acoustic representations are transformed by a DNN into another type of frame-level features. The DNN output features have variable length, which is determined by the time duration of input utterance. A method of aggregation is applied to convert the variable-length DNN features into a fixed-length embedding that represents speaker's characteristics as reflected in the utterance. Given a test utterance and a reference utterance, speaker verification is performed based on similarity measure between their embeddings. $Average\ Pooling$, i.e., taking simple time average of frame-level DNN features, is an intuitive approach to feature aggregation \cite{Chung18b}. $Statistics\ Pooling$ \cite{snyder2017deep} computes the mean and standard deviation of frame-level features as the utterance-level representation. In these methods, features from different parts of the utterance are assumed to be equally important, and temporal relation between the features is not considered. In \cite{heigold2016end}, a recurrent neural network (RNN) is used to capture temporal dependency and derive utterance-level embedding for SV.

%Attention mechanism in vision and speech recognition...
Attention mechanism in DNN has been shown effective in various application areas \cite{chorowski2015attention,zeyer2018improved,vaswani2017attention,wang2018non}. In simple terms, attention to selected parts of a feature is realized through a method of determining and imposing heavier weights, so as to make these parts more salient and play a more important role in the intended task. In the case of SV, attention mechanism can be implemented in the process of aggregating frame-level features with learned attention weights \cite{bhattacharya2017deep,okabe2018attentive}. This approach showed better performance than $Average$ or $Statistics\ Pooling$, confirming that speaker-relevant information is not evenly distributed in an utterance. Typically the attention weights used to compute the embedding for an input utterance are derived from this utterance itself. This is known as self-attention. As the SV process involves two input utterances, the information from both utterances could be exploited to improve the attention mechanism. This idea of collaborative attention was applied in video-based person re-identification\cite{zhang2019scan} and text-dependent SV\cite{zhang2019seq2seq}. In this paper, the use of $Dual\ Attention$ mechanism is proposed for combining frame-level DNN features in text-independent SV. The attention model comprises two component modules:
\begin{itemize}
\item $Self$-$attention$: the attention weights for each of the two utterances are computed from DNN features of the utterance itself;
\item $Mutual$-$attention$: DNN features of the two utterances collaborate with each other to generate the attention weights
\end{itemize}

The utterance-level embeddings generated by the $Dual\ Attention$ model are passed to a binary classifier to determine whether the two utterances are from the same speaker or not. The classifier output can be regarded as a similarity score. It is further combined with a cosine distance to produce the final similarity score. The proposed $Dual\ Attention\ Network$ (abbreviated as D-att Net) is trained end-to-end.

\section{The proposed model}
D-att Net contains three major parts: (1) a backbone network for extracting frame-level DNN features from an input utterance; (2) an attention network that aggregates features from the backbone to generate utterance-level embeddings; (3) a decision module that fuses the similarity score from a binary classifier with the cosine distance.

\subsection{The backbone network}
The backbone network follows the ResNet\cite{he2016deep} structure and takes spectrogram of a speech segment as single channel input. The spectrogram has the size of $T\times F$, where $T$ is the time dimension (number of frames) and $F$ is the frequency dimension (number of Mel-filterbanks). The first convolution (Conv) layer of the ResNet is replaced by a pre-processing block in the proposed model. This block begins with a batch normalization (BN) layer, and the normalized features are subsequently processed via two separate streams. The first stream involves a Conv layer. All input patches across the input spectrogram are processed with the same convolution operation, regardless of the patches' time and frequency locations. As a consequence, two local patches that show the same pattern but are from different frequency regions would become non-distinguishable. This limitation is addressed by another Conv layer in the second stream, in which convolution is applied on each frequency bin of the input, with kernel size $1\times1$ and kernel depth $F$. The number of filters in this Conv layer is equal to the frequency dimension of input spectrogram, such that the output feature has the same size as the input. Outputs from the two streams are concatenated along the channel dimension for subsequent ResNet processing.

\begin{table}
        \caption{The backbone network. A batch normalization layer and a ReLU layer are added following each convolution and fully connected layer, except for $FC2$. $T$ and $F$ are the time and frequency dimensions of the input spectrogram. $num_{f}$ is the dimension of the backbone network's output features, and $num_{ID}$ is the number of speakers in the SV task.}
        \label{tab:Backbone_structure}
        \small{
        \centering
        \scalebox{0.83}{
        \begin{tabular}{l|c|c|c}
        \hline 
        \textbf{Block} & \multicolumn{2}{c|}{\textbf{Structure}} & \textbf{Output size} \\
        \hline
        \hline
         & Stream1 &  Stream2 & Height/Width/Channel \\
        \cline{2-2}  \cline{3-3}  \cline{4-4}
         & \multicolumn{2}{|c|}{BN} & {$T\times F \times1$} \\
         \cline{2-2}  \cline{3-3}  \cline{4-4}
         & &  & Stream1: \\ 
        Pre- & Conv2D, &  Conv2D, & {$T\times F \times16$}\\ 
        processing & $7\times7$, stride 1 & $1\times1$, stride 1 & Stream2:\\
        Block & & &{$T\times F \times1$}\\
        \cline{2-2}  \cline{3-3}  \cline{4-4} 
         & \multicolumn{2}{|c|}{Concatenate} & {$T \times F \times 17$}\\
        \cline{2-2}  \cline{3-3}  \cline{4-4} 
         & \multicolumn{2}{|c|}{Conv2D, $1\times1$, stride 1}& {$T\times F \times64$} \\
         \hline 
         & \multicolumn{2}{|c|}{MaxPool1, $3\times3$, stride 2}& {$T/2 \times F/2 \times 64$} \\
        \cline{2-2}  \cline{3-3}  \cline{4-4}
         & \multicolumn{2}{|c|}{ResNet Block1}& {$T/2 \times F/2 \times 64$} \\
        \cline{2-2}  \cline{3-3}  \cline{4-4}
        ResNet & \multicolumn{2}{|c|}{ResNet Block2}& {$T/4 \times F/4 \times 128$} \\
        \cline{2-2}  \cline{3-3}  \cline{4-4}
        Backbone & \multicolumn{2}{|c|}{MaxPool2, $3\times1$, stride $2\times1$}& {$T/8 \times F/4 \times 128$} \\
        \cline{2-2}  \cline{3-3}  \cline{4-4}
         & \multicolumn{2}{|c|}{ResNet Block3}& {$T/16 \times F/8 \times 256$} \\
        \cline{2-2}  \cline{3-3}  \cline{4-4}
         & \multicolumn{2}{|c|}{ResNet Block4}& {$T/32 \times F/16 \times 512$} \\
        \hline 
         & \multicolumn{2}{|c|}{AvgPool1, $1\times F/16$, stride 1}& {$T/32 \times 512$} \\
        \cline{2-2}  \cline{3-3}  \cline{4-4}
        Post- & \multicolumn{2}{|c|}{FC1}& {$T/32 \times num_{f}$} \\
        \cline{2-2}  \cline{3-3}  \cline{4-4}
        processing & \multicolumn{2}{|c|}{AvgPool2, $T/32$, stride 1}& {$num_{f}$} \\
        \cline{2-2}  \cline{3-3}  \cline{4-4}
        Block & \multicolumn{2}{|c|}{FC2} & {$num_{ID}$} \\
        \cline{2-2}  \cline{3-3}  \cline{4-4}
         & \multicolumn{2}{|c|}{Cross-Entropy Loss}& {-} \\
        \hline 
        \end{tabular}
        }}
\end{table}

The details of the backbone network are shown as Table~\ref{tab:Backbone_structure}. A sequence of DNN features are extracted from $Avg Pool1$ with the size of $T/32 \times 512$. Hereafter we use $T' = T/32$ to denote the compressed time length. The fully connected layer $FC1$ produces the frame-level DNN features with length $T'$ and feature dimension $num_{f}$. The frame-level features are averaged along the time dimension and passed to $FC2$ for generating speaker ID.

\subsection{The attention network}
In a typical application scenario of SV, the test utterance is a few seconds long, containing a number of phonemes. Some parts of the utterance may be produced with more speaker-specific characteristics and some with less. With the attention network, the utterance-level embedding is obtained as an attention-weighted sum of the $T'$ frame-level features.

\begin{figure}[t]
  \centering
  \includegraphics[width=\linewidth]{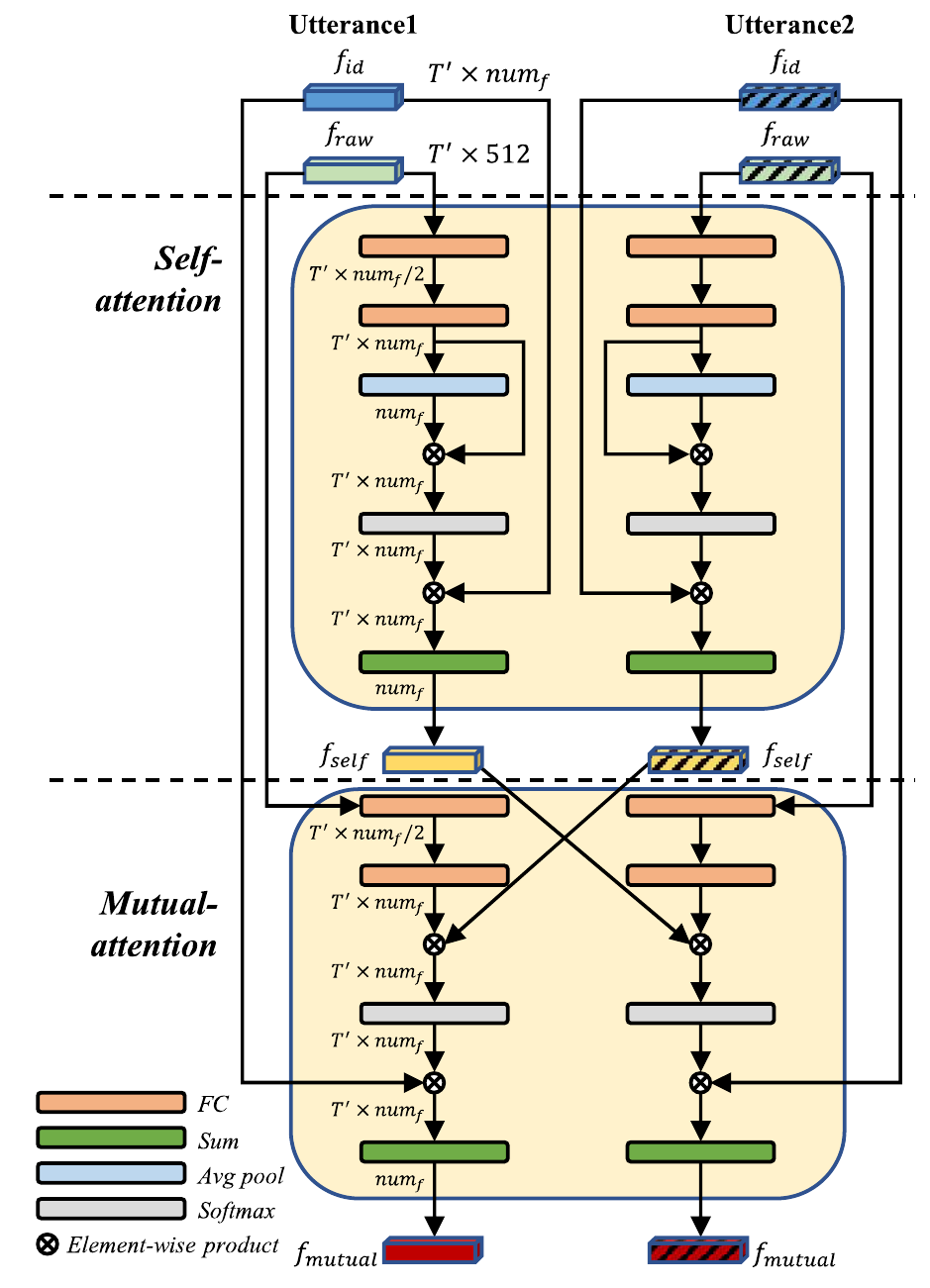}
  \caption{Structure of the attention network. Two utterances' features are differentiated by textures. This structure is symmetric, and the layers and parameters are shared for the two utterances. For the element-wise product of tensors with different sizes, we first duplicate the smaller one multiple times to match the size of the larger one.}
  \label{fig:attention}
\end{figure}

As shown in Figure~\ref{fig:attention}, the attention network for SV takes in two input utterances. They are named as ``Utterance 1'' and ``Utterance 2'', without explicitly specifying the test utterance and the reference utterance. For each utterance, there are two input representations $f_{raw}$ and $f_{id}$, which are generated from $Avg Pool1$ and $FC1$ of the backbone network respectively (see Table~\ref{tab:Backbone_structure}). The features for the two utterances are illustrated by cuboids with different textures.

\subsubsection{Self-Attention}
Let $\{f_{raw}^{i}\}^{T'}_{i=1}$ be the set of $T'$ frame-level features (from $Avg Pool1$ of the backbone network). They are transformed by two FC layers to produce $\{f_{att}^{i}\}^{T'}_{i=1}$, which are further converted into the self-attention weight matrix $W_{self}$ as,
\begin{equation}
  W_{self} = Softmax(\{f_{att}^{i}\}^{T'}_{i=1}\otimes\frac{1}{T'}\sum_{i=1}^{T'}f_{att}^{i})
  \label{eq_self}
\end{equation}
where $\otimes$ denotes element-wise product. $\{f_{att}^{i}\}^{T'}_{i=1}$ is first scaled by its time average. The $Softmax$ function is applied to normalize the scaled values across different time frames. The size of $W_{self}$ is $T' \times num_f$. Each channel of the DNN features is assigned a distinct attention weight on each of the $T'$ frames. The self-attention weighted feature $f_{self}$ for the respective utterance is obtained by summing up the elements of $W_{self} \otimes f_{id}$ along the time dimension.

%The frame-level attention weights $\{f_{att}^{i}\}^{T'}_{i=1}$ is averaged over time dimension to obtain a global attention weight. The element-wise product of $\{f_{att}^{i}\}^{T'}_{i=1}$ and the global attention weight is processed by $Softmax$ to normalize the weight values across different time frames. $W_{self}$ has size $T' \times num_f$, representing each channel of the DNN features has its own attention weights on $T'$ frames. The output feature $f_{self}$ is obtained by the sum of $W_{self} \otimes f_{id}$ on time dimension, giving a utterance-level feature representation of the input utterance.

\subsubsection{Mutual-Attention}
The motivation of incorporating mutual attention is to leverage mutually discriminative parts of the two input utterances. Similar to the self-attention module, $\{f_{att}^{i}\}^{T'}_{i=1}$ with size $T' \times num_{f}$ is generated by two FC layers. For each of the two utterances, the mutual-attention weight matrix $W_{mutual}$ is obtained by using the $f_{self}$ feature from the other utterance. As an example, for Utterance 1 we have 
\begin{equation}
  (W_{mutual})^{(1)} = Softmax((\{f_{att}^{i}\}^{T'}_{i=1})^{(1)} \otimes (f_{self})^{(2)})
  \label{eq_mutual}
\end{equation}
where the superscript labels ``(1)” and ``(2)” are used to denote Utterance 1 and Utterance 2 respectively. Summing up the elements in $(W_{mutual})^{(1)} \otimes (f_{id})^{(1)}$ along the time dimension gives the mutual-attention weighted feature $f_{mutual}$ for Utterance 1.

\subsection{Similarity estimation}
In the proposed model, similarity between two input utterances is measured with the cosine distance and the output score of a binary classifier. The cosine distance $score_{cos}$ is computed on the output of $Avg Pool2$ of the backbone network.
The binary classifier uses the Sigmoid function to produce a similarity score based on the attention-weighted features, 
\begin{equation}
  \begin{split}
  score_{binary} = Sigmoid&(FC(BN(((f_{self})^{(1)}-(f_{self})^{(2)})\\
  &\otimes((f_{mutual})^{(1)}-(f_{mutual})^{(2)}))))
  \end{split}
  \label{eq_binary_score}
\end{equation}
where $(f_{self})^{(1)}$, $(f_{mutual})^{(1)}$ denote the features from one of the utterances, and $(f_{self})^{(2)}$, $(f_{mutual})^{(2)}$ from the other one. The binary classifier is trained toward output value of ``1" if the two utterances are from the same speaker and ``0" otherwise.

The cosine distance score $score_{cos}$ and the classifier output score $score_{binary}$ are normalized separately using the global mean and standard deviation obtained from a large number of utterance pairs randomly sampled from training data. The combined overall score $score_{all}$ is equal to the average of normalized cosine distance score and binary classifier score.

%\begin{equation}
%  score_{all} = \frac{Norm(score_{cos}) + Norm(score_{binary})}{2}
%  \label{eq_final_score}
%\end{equation}

\section{Experiments}

\subsection{Dataset}
The speech databases used in this study are Voxceleb1 and Voxceleb2 \cite{Nagrani17,Chung18b}. Voxceleb1 contains $1,211$ speakers in the development set and $40$ speakers in the test set. Voxceleb2 has $5,994$ speakers in the development set. The development sets of Voxceleb1 and Voxceleb2 are jointly utilized for model training in the following experiments. Thus the training data comprises about 1.2 million utterances from $7,205$ speakers ($num_{ID}=7,205$). For performance evaluation, $37,720$ pairs of utterances are formed from $4,874$ utterances in Voxceleb1 test set.

%In the test set of Voxceleb1, $4,874$ utterances are generated by $40$ speakers and $37,720$ pairs are formed from these utterances for verification. The test set has no overlap with the speakers in the development set.
%In the test set of Voxceleb1, $37,720$ pairs are formed from $4,874$ utterances for verification. 

The audio signals at sampling rate of 16 kHz are divided into short-time frames of 25 ms with 10 ms frame shift. Each frame is represented by 512-point DFT spectrum. 64-dimension log Mel-filterbank (FBank) coefficients are calculated from the short-time spectrum and used as the input of the backbone network. The acoustic signal processing functions are implemented with the Librosa library \cite{mcfee2015librosa}.

\subsection{Training details}
In the training process, a three-second segment is randomly cropped from each utterance. This gives an input of the size $300 \times 64$ for the backbone network. In all experiments, the dimension of the DNN output features, $num_{f}$, is fixed at $256$.

% The loss function for the speaker identification task, denoted by $loss_{id}$ is defined as the cross-entropy loss on the output of $FC2$ of the backbone network. Each step of training involves $64$ randomly selected speakers in the training set. Two utterances are provided by each of the speakers. The $128$ utterances are divided into two groups, referred to as Group 1 and Group 2 respectively, each containing $64$ utterances. One utterance from Group 1 and one from Group 2 form an input pair for training, giving a total of $64\times64=4,096$ training pairs. As shown in Figure 1, the DNN output features generated from ``Utterance 1” and ``Utterance 2” in a training pair are processed by the attention network. The attention-weighted features are passed to the binary classifier. The cross-entropy loss evaluated at the classifier output is denoted as $loss_{binary}$.

The loss function for the speaker identification task, denoted by $loss_{id}$ is defined as the cross-entropy loss on the output of $FC2$ of the backbone network. Each step of training involves $64$ randomly selected speakers in the training set. Two utterances are provided by each of the speakers and put into two groups, referred as Group 1 and Group 2 respectively. Therefore, there are 128 utterances in one batch (64 utterances for each group). One utterance from Group 1 and one from Group 2 form an input pair for training, giving a total of $64\times64=4,096$ training pairs. As shown in Figure 1, the DNN output features generated from ``Utterance 1” and ``Utterance 2” in a training pair are processed by the attention network. The attention-weighted features are passed to the binary classifier for $score_{binary}$ calculation. The cross-entropy loss evaluated at the binary classifier output is denoted as $loss_{binary}$.

The final loss is given by the sum of $loss_{id}$ and $loss_{binary}$ as follow:
\begin{equation}
  loss_{all} = loss_{id} + \lambda loss_{binary}
  \label{eq_final_loss}
\end{equation}
where $\lambda$ is an empirically determined parameter to control the weight of $loss_{binary}$. Different values of $\lambda$ are evaluated in the experiments.

Model training was implemented with PyTorch\cite{paszke2019pytorch} and two GPUs. The optimizer used is Stochastic Gradient Descent, with $0.9$ momentum and $0.001$ weight decay. The initial learning rate is $0.1$ for the backbone, $0.01$ for the attention network and the binary classifier. The learning rate decreases following a half cosine shape\cite{loshchilov2016sgdr}. To avoid over-fitting, a dropout layer is included in the binary classifier, with a dropout rate of $0.5$. All networks are trained end-to-end for $20$ epochs.

\subsection{Performance evaluation}
Each step of performance evaluation involves a pair of test utterances, which could be from the same speaker or two different speakers. Each utterance is divided into segments of $5$ seconds long, with $4$ seconds overlap between two neighboring segments. If an utterance is shorter than $5$ seconds, the means of its frame-level Fbank coefficients are appended at the end of the FBank sequences so as to equalize the size of input representation to a $5$-second long segment. Similarity estimation is done on all pairing combinations of segments from the two utterances. For example, if utterance 1 contains $X$ segments and utterance 2 has $Y$ segments, there would be $X \times Y$ cosine distances computed. The average of these distances gives $score_{cos}$ for the two utterances. $score_{binary}$ is obtained in a similar way by averaging the binary classifier output scores of the $X \times Y$ segment pairs. 

\section{Results}

\subsection{Baseline}
The ResNet18 shown in Table~\ref{tab:Backbone_structure} is regarded as the baseline model, using only the cosine distance $score_{cos}$ for similarity estimation. The baseline system's performance is shown in Table~\ref{tab:EER_baseline_result1} and it achieves EER$=2.6\%$, which noticeably outperforms the result on standard ResNet18 structure. To evaluate the effect of binary classifier on performance gain, a binary classifier is added into the baseline. The output score is calculated by Equation~\ref{eq_binary_score}, with $f_{self}$ and $f_{mutual}$ being replaced by the output of $Avg Pool2$. The binary classifier is jointly trained with the backbone network, and the classifier output score is fused with $score_{cos}$ as described in similarity estimation. There is no improvement observed, as compared with $score_{cos}$. This suggests that the binary classifier could not yield better results using only the features from the backbone.

%By using $AM\textit{-}Softmax$ or deeper structure ResNet34 we can see the baseline's performance improve slightly. 

\begin{table}[t]
  \caption{Performances of our models. All results are evaluated on Voxceleb1 test set.}
  \label{tab:EER_baseline_result1}
  \centering
  \begin{tabular}{lc}
    \toprule
    \textbf{Model}                       & \textbf{EER(\%)}  \\
    \midrule
    ResNet18(Standard)                   & 2.91              \\
    ResNet18(Ours)                       & 2.60              \\
    ResNet18(Ours)+Binary                & 2.59              \\
    \hline
    Dual Attention Net, $\lambda=0.5$    & 2.53              \\
    Dual Attention Net, $\lambda=1$      & \textbf{2.49}     \\
    Dual Attention Net, $\lambda=2$      & 2.55              \\
    \bottomrule
  \end{tabular}
\end{table}

\subsection{D-att Net}
The proposed D-att network is evaluated in three training settings with $\lambda$ being $0.5$, $1$ and $2$. The results of D-att Net in Table~\ref{tab:EER_baseline_result1} are evaluated on $score_{all}$. The best performance EER$=2.49\%$ is achieved with $\lambda=1$. The performance drops when we decrease or increase $\lambda$ in the experiment. Decreasing $\lambda$ may weaken the learning ability of attention mechanism and the binary classifier, while increasing $\lambda$ may push the binary classifier to over-fitting.

\begin{table}[t]
  \caption{Performances of our models. All results are evaluated on Voxceleb1 test set.            $\lambda$ equals 1.}
  \label{tab:EER_baseline_result2}
  \centering
  \begin{tabular}{llc}
    \toprule
    \textbf{Model} &                                 & \textbf{EER(\%)}    \\
    \midrule
    \multicolumn{2}{l}{ResNet18(Ours)+Softmax}       & 2.60                \\
    \multicolumn{2}{l}{ResNet34(Ours)+Softmax}       & 2.35                \\
    \multicolumn{2}{l}{ResNet18(Ours)+AM-Softmax}    & 2.49                \\
    \multicolumn{2}{l}{ResNet34(Ours)+AM-Softmax}    & 2.16                \\
    \hline
    \multirow{3}{*}{\shortstack[l]{Dual Attention Net, \\ Res18,Softmax}}    
      & $score_{cos}$         & 2.60           \\
      & $score_{binary}$      & 2.71           \\
      & $score_{all}$         & 2.49           \\
    \hline
    \multirow{3}{*}{\shortstack[l]{Dual Attention Net, \\ Res34,Softmax}}    
      & $score_{cos}$         & 2.31           \\
      & $score_{binary}$      & 2.43           \\
      & $score_{all}$         & 2.24           \\
    \hline
    \multirow{3}{*}{\shortstack[l]{Dual Attention Net, \\ Res18,AM-Softmax}}  
      & $score_{cos}$         & 2.18           \\
      & $score_{binary}$      & 2.49           \\
      & $score_{all}$         & \textbf{1.88}  \\
    \hline
    \multirow{3}{*}{\shortstack[l]{Dual Attention Net, \\ Res34,AM-Softmax}}  
      & $score_{cos}$         & 1.84           \\
      & $score_{binary}$      & 2.12           \\
      & $score_{all}$         & \textbf{1.60}  \\
    \bottomrule
  \end{tabular}
\end{table}
 
% $score_{cos}$, $score_{binary}$, $score_{all}$ are involved in the similarity estimation in D-att Net. We evaluate the EER of these three scores on the ResNet18 and ResNet34 backbones. The results are shown in Table~\ref{tab:EER_baseline_result2}. $score_{all}$ outperforms the baselines' results and yields better performance than the other two scores, indicating $score_{cos}$ and $score_{binary}$ are complementary. Notably, $score_{cos}$ in D-att Net achieves better result than the baseline under the same backbone structure, which demonstrates our attention model and binary classifier contribute to feature learning in the backbone.
 
 Recently, $Additive\ Margin\ Softmax$($AM\textit{-}Softmax$) was investigated by \cite{wang2018additive,wang2018cosface} in face recognition to replace $Softmax$ in the cross-entropy loss. $AM\textit{-}Softmax$ was also applied to SV\cite{li2019boundary,liu2019large} with some modifications. It's calculated as: 
\begin{equation}
  \hat{y_{i}} = \frac{e^{s(cos\theta_{y_i}-m)}}{e^{s(cos\theta_{y_i}-m)}+\sum_{j\neq y_i}^{num_{id}}e^{s(cos\theta_{y_j})}}
  \label{eq_AM_softmax}
\end{equation}
where $cos\theta_{y_i}$ is the cosine distance between input feature and classification weights $w_i$ in $FC2$. $s$ is a hyperparameter for scaling and $m$ controls the cosine margin. $AM\textit{-}Softmax$ is applied with $s=30$ and $m=0.2$ in our experiments and the results are shown in Table~\ref{tab:EER_baseline_result2}. Utilizing $Dual\ Attention$, $AM\textit{-}Softmax$ achieves larger performance improvement than $Softmax$ and decreases the baseline's EER by around $20\%$.

$score_{cos}$, $score_{binary}$, $score_{all}$ are involved in the similarity estimation in D-att Net, and the EER of these three scores are reported in Table~\ref{tab:EER_baseline_result2}. $score_{all}$ outperforms the baseline's result and yields better performance than the other two scores, indicating $score_{cos}$ and $score_{binary}$ are complementary. Notably, $score_{cos}$ in D-att Net achieves lower EER than the baseline under the same backbone structure, which demonstrates our attention model and binary classifier contribute to feature learning in the backbone network.

\begin{table}[t]
  \caption{Performances of different models. Soft. is short for Softmax. SP stands for Statistics Pooling. Models marked with * use only Voxceleb2 development set for training. Models marked with ** use extra data MUSAN\cite{snyder2015musan} and RIR\cite{ko2017study} for data augmentation. All models are evaluated on Voxceleb1 test set.}
  \label{tab:EER_model_result}
  \centering
  \begin{tabular}{llc}
    \toprule
    \textbf{Model}   &\textbf{Aggregation}   & \textbf{EER(\%)}          \\
    \midrule
    TDNN+PLDA\cite{snyder2018x}**            & SP        & 3.10          \\
    Thin-ResNet34+binary\cite{Nagrani19}*    & GhostVLAD & 2.87          \\
    DDB+Gate+cosine\cite{jiang2019effective} & SP        & 2.91          \\
    DDB+Gate+PLDA\cite{jiang2019effective}   & SP        & 2.31          \\
    Large Margin-Soft.\cite{liu2019large}**  & SP        & 2.00          \\
    \hline
    D-att Net(Res18,AM-Soft.)                & Dual-att  & \textbf{1.88} \\
    D-att Net(Res34,AM-Soft.)                & Dual-att  & \textbf{1.60} \\
    \bottomrule
  \end{tabular}
\end{table}

\subsection{Comparison with other models}
The proposed D-att Net is compared with state-of-the-art systems in Table~\ref{tab:EER_model_result}. In \cite{snyder2018x}, data augmentation was applied with TDNN and it boosted the performance in a large step. \cite{Nagrani19} applied $GhostVLAD$ for aggregating frame-level features and utilized a binary classifier to estimate similarity between utterances. By adding dilated dense block(DDB) and gating mechanism into TDNN, \cite{jiang2019effective} proposed a modified network structure. \cite{liu2019large} modified $AM\textit{-}Softmax$ and achieved significant result with the help of data augmentation. Our proposed D-att Net models yield the best performance among all these works, giving $1.88\%$ EER on the ResNet18 structure and $1.6\%$ EER on the ResNet34.

\section{Conclusion and future work}
In this paper, we present a $Dual\ Attention$ structure with modified ResNet backbone for text-independent speaker verification. The frame-level DNN features are extracted by the backbone network and aggregated by the proposed attention model. In the attention model, self-attention and mutual-attention are involved to combine two input utterances' information in the calculation of attention weights and generation of utterance-level embedding. The embeddings are utilized to predict the similarity between two utterances by fusing a binary classifier output score with cosine distance. The full model, D-att Net, achieves state-of-the-art performance on Voxceleb1 test set in our experiments, which demonstrates the effectiveness of our proposed network.

In the future we will evaluate more backbone structures and combine PLDA or other scoring methods in our work.

% \section{Acknowledgements}

\bibliographystyle{IEEEtran}

\bibliography{mybib}

\begin{thebibliography}{10}
\providecommand{\url}[1]{#1}
\csname url@samestyle\endcsname
\providecommand{\newblock}{\relax}
\providecommand{\bibinfo}[2]{#2}
\providecommand{\BIBentrySTDinterwordspacing}{\spaceskip=0pt\relax}
\providecommand{\BIBentryALTinterwordstretchfactor}{4}
\providecommand{\BIBentryALTinterwordspacing}{\spaceskip=\fontdimen2\font plus
\BIBentryALTinterwordstretchfactor\fontdimen3\font minus
  \fontdimen4\font\relax}
\providecommand{\BIBforeignlanguage}[2]{{%
\expandafter\ifx\csname l@#1\endcsname\relax
\typeout{** WARNING: IEEEtran.bst: No hyphenation pattern has been}%
\typeout{** loaded for the language `#1'. Using the pattern for}%
\typeout{** the default language instead.}%
\else
\language=\csname l@#1\endcsname
\fi
#2}}
\providecommand{\BIBdecl}{\relax}
\BIBdecl

\bibitem{variani2014deep}
E.~Variani, X.~Lei, E.~McDermott, I.~L. Moreno, and J.~Gonzalez-Dominguez,
  ``Deep neural networks for small footprint text-dependent speaker
  verification,'' in \emph{2014 IEEE International Conference on Acoustics,
  Speech and Signal Processing (ICASSP)}.\hskip 1em plus 0.5em minus
  0.4em\relax IEEE, 2014, pp. 4052--4056.

\bibitem{campbell1997speaker}
J.~P. Campbell, ``Speaker recognition: A tutorial,'' \emph{Proceedings of the
  IEEE}, vol.~85, no.~9, pp. 1437--1462, 1997.

\bibitem{snyder2018x}
D.~Snyder, D.~Garcia-Romero, G.~Sell, D.~Povey, and S.~Khudanpur, ``X-vectors:
  Robust dnn embeddings for speaker recognition,'' in \emph{2018 IEEE
  International Conference on Acoustics, Speech and Signal Processing
  (ICASSP)}.\hskip 1em plus 0.5em minus 0.4em\relax IEEE, 2018, pp. 5329--5333.

\bibitem{Chung18b}
J.~S. Chung, A.~Nagrani, and A.~Zisserman, ``Voxceleb2: Deep speaker
  recognition,'' in \emph{INTERSPEECH}, 2018.

\bibitem{snyder2017deep}
D.~Snyder, D.~Garcia-Romero, D.~Povey, and S.~Khudanpur, ``Deep neural network
  embeddings for text-independent speaker verification.'' in
  \emph{Interspeech}, 2017, pp. 999--1003.

\bibitem{heigold2016end}
G.~Heigold, I.~Moreno, S.~Bengio, and N.~Shazeer, ``End-to-end text-dependent
  speaker verification,'' in \emph{2016 IEEE International Conference on
  Acoustics, Speech and Signal Processing (ICASSP)}.\hskip 1em plus 0.5em minus
  0.4em\relax IEEE, 2016, pp. 5115--5119.

\bibitem{chorowski2015attention}
J.~K. Chorowski, D.~Bahdanau, D.~Serdyuk, K.~Cho, and Y.~Bengio,
  ``Attention-based models for speech recognition,'' in \emph{Advances in
  neural information processing systems}, 2015, pp. 577--585.

\bibitem{zeyer2018improved}
A.~Zeyer, K.~Irie, R.~Schl{\"u}ter, and H.~Ney, ``Improved training of
  end-to-end attention models for speech recognition,'' \emph{arXiv preprint
  arXiv:1805.03294}, 2018.

\bibitem{vaswani2017attention}
A.~Vaswani, N.~Shazeer, N.~Parmar, J.~Uszkoreit, L.~Jones, A.~N. Gomez,
  {\L}.~Kaiser, and I.~Polosukhin, ``Attention is all you need,'' in
  \emph{Advances in neural information processing systems}, 2017, pp.
  5998--6008.

\bibitem{wang2018non}
X.~Wang, R.~Girshick, A.~Gupta, and K.~He, ``Non-local neural networks,'' in
  \emph{Proceedings of the IEEE conference on computer vision and pattern
  recognition}, 2018, pp. 7794--7803.

\bibitem{bhattacharya2017deep}
G.~Bhattacharya, M.~J. Alam, and P.~Kenny, ``Deep speaker embeddings for
  short-duration speaker verification.'' in \emph{Interspeech}, 2017, pp.
  1517--1521.

\bibitem{okabe2018attentive}
K.~Okabe, T.~Koshinaka, and K.~Shinoda, ``Attentive statistics pooling for deep
  speaker embedding,'' \emph{arXiv preprint arXiv:1803.10963}, 2018.

\bibitem{zhang2019scan}
R.~Zhang, J.~Li, H.~Sun, Y.~Ge, P.~Luo, X.~Wang, and L.~Lin, ``Scan:
  Self-and-collaborative attention network for video person
  re-identification,'' \emph{IEEE Transactions on Image Processing}, vol.~28,
  no.~10, pp. 4870--4882, 2019.

\bibitem{zhang2019seq2seq}
Y.~Zhang, M.~Yu, N.~Li, C.~Yu, J.~Cui, and D.~Yu, ``Seq2seq attentional siamese
  neural networks for text-dependent speaker verification,'' in \emph{ICASSP
  2019-2019 IEEE International Conference on Acoustics, Speech and Signal
  Processing (ICASSP)}.\hskip 1em plus 0.5em minus 0.4em\relax IEEE, 2019, pp.
  6131--6135.

\bibitem{he2016deep}
K.~He, X.~Zhang, S.~Ren, and J.~Sun, ``Deep residual learning for image
  recognition,'' in \emph{Proceedings of the IEEE conference on computer vision
  and pattern recognition}, 2016, pp. 770--778.

\bibitem{Nagrani17}
A.~Nagrani, J.~S. Chung, and A.~Zisserman, ``Voxceleb: a large-scale speaker
  identification dataset,'' in \emph{INTERSPEECH}, 2017.

\bibitem{mcfee2015librosa}
B.~McFee, C.~Raffel, D.~Liang, D.~P. Ellis, M.~McVicar, E.~Battenberg, and
  O.~Nieto, ``librosa: Audio and music signal analysis in python,'' in
  \emph{Proceedings of the 14th python in science conference}, vol.~8, 2015.

\bibitem{paszke2019pytorch}
A.~Paszke, S.~Gross, F.~Massa, A.~Lerer, J.~Bradbury, G.~Chanan, T.~Killeen,
  Z.~Lin, N.~Gimelshein, L.~Antiga \emph{et~al.}, ``Pytorch: An imperative
  style, high-performance deep learning library,'' in \emph{Advances in Neural
  Information Processing Systems}, 2019, pp. 8024--8035.

\bibitem{loshchilov2016sgdr}
I.~Loshchilov and F.~Hutter, ``Sgdr: Stochastic gradient descent with warm
  restarts,'' \emph{arXiv preprint arXiv:1608.03983}, 2016.

\bibitem{wang2018additive}
F.~Wang, J.~Cheng, W.~Liu, and H.~Liu, ``Additive margin softmax for face
  verification,'' \emph{IEEE Signal Processing Letters}, vol.~25, no.~7, pp.
  926--930, 2018.

\bibitem{wang2018cosface}
H.~Wang, Y.~Wang, Z.~Zhou, X.~Ji, D.~Gong, J.~Zhou, Z.~Li, and W.~Liu,
  ``Cosface: Large margin cosine loss for deep face recognition,'' in
  \emph{Proceedings of the IEEE Conference on Computer Vision and Pattern
  Recognition}, 2018, pp. 5265--5274.

\bibitem{li2019boundary}
R.~Li, N.~Li, D.~Tuo, M.~Yu, D.~Su, and D.~Yu, ``Boundary discriminative large
  margin cosine loss for text-independent speaker verification,'' in
  \emph{ICASSP 2019-2019 IEEE International Conference on Acoustics, Speech and
  Signal Processing (ICASSP)}.\hskip 1em plus 0.5em minus 0.4em\relax IEEE,
  2019, pp. 6321--6325.

\bibitem{liu2019large}
Y.~Liu, L.~He, and J.~Liu, ``Large margin softmax loss for speaker
  verification,'' \emph{arXiv preprint arXiv:1904.03479}, 2019.

\bibitem{snyder2015musan}
D.~Snyder, G.~Chen, and D.~Povey, ``Musan: A music, speech, and noise corpus,''
  \emph{arXiv preprint arXiv:1510.08484}, 2015.

\bibitem{ko2017study}
T.~Ko, V.~Peddinti, D.~Povey, M.~L. Seltzer, and S.~Khudanpur, ``A study on
  data augmentation of reverberant speech for robust speech recognition,'' in
  \emph{2017 IEEE International Conference on Acoustics, Speech and Signal
  Processing (ICASSP)}.\hskip 1em plus 0.5em minus 0.4em\relax IEEE, 2017, pp.
  5220--5224.

\bibitem{Nagrani19}
A.~Nagrani, J.~S. Chung, W.~Xie, and A.~Zisserman, ``Voxceleb: Large-scale
  speaker verification in the wild,'' \emph{Computer Science and Language},
  2019.

\bibitem{jiang2019effective}
Y.~Jiang, Y.~Song, I.~McLoughlin, Z.~Gao, and L.~Dai, ``An effective deep
  embedding learning architecture for speaker verification,'' \emph{Proc.
  Interspeech 2019}, pp. 4040--4044, 2019.

\end{thebibliography}

% \begin{thebibliography}{9}
% \bibitem[1]{Davis80-COP}
%   S.\ B.\ Davis and P.\ Mermelstein,
%   ``Comparison of parametric representation for monosyllabic word recognition in continuously spoken sentences,''
%   \textit{IEEE Transactions on Acoustics, Speech and Signal Processing}, vol.~28, no.~4, pp.~357--366, 1980.
% \bibitem[2]{Rabiner89-ATO}
%   L.\ R.\ Rabiner,
%   ``A tutorial on hidden Markov models and selected applications in speech recognition,''
%   \textit{Proceedings of the IEEE}, vol.~77, no.~2, pp.~257-286, 1989.
% \bibitem[3]{Hastie09-TEO}
%   T.\ Hastie, R.\ Tibshirani, and J.\ Friedman,
%   \textit{The Elements of Statistical Learning -- Data Mining, Inference, and Prediction}.
%   New York: Springer, 2009.
% \bibitem[4]{YourName17-XXX}
%   F.\ Lastname1, F.\ Lastname2, and F.\ Lastname3,
%   ``Title of your INTERSPEECH 2020 publication,''
%   in \textit{Interspeech 2020 -- 20\textsuperscript{th} Annual Conference of the International Speech Communication Association, September 15-19, Graz, Austria, Proceedings, Proceedings}, 2020, pp.~100--104.
% \end{thebibliography}

\end{document}